Prediction of activation energy barrier of island diffusion processes using data-driven approaches

Shree Ram Acharya[1] and Talat S. Rahman[1, 2*]

[1]Department of Physics, University of Central Florida, Orlando, FL 32816, USA

[2]Donostia International Physics Center, Donostia-San Sebastian, 20018, Spain

We present models for prediction of activation energy barrier of diffusion process of adatom (1-4) islands obtained by using data-driven techniques. A set of easily accessible features, geometric and energetic, that are extracted by analyzing the variation of the energy barriers of a large number of processes on homo-epitaxial metallic systems of Cu, Ni, Pd, and Ag are used along with the calculated activation energy barriers to train and test linear and non-linear statistical models. A multivariate linear regression model trained with energy barriers for Cu, Pd, and Ag systems explains 92% of the variation of energy barriers of the Ni system, whereas the non-linear model using artificial neural network slightly enhances the success to 93%. Next mode of calculation that uses barriers of all four systems in training, predicts barriers of randomly picked processes of those systems with significantly high correlation coefficient: 94.4% in linear regression model and 97.7% in artificial neural network model. Calculated kinetics parameters such as the type of frequently executed processes and effective energy barrier for Ni dimer and trimer diffusion on the Ni(111) surface obtained from KMC simulation using the predicted (data-enabled) energy barriers are in close agreement with those obtained by using energy barriers calculated from interatomic interaction potential.

I. INTRODUCTION

The development of computational approaches which enable long time simulation of atomic system that reveal rare events responsible for system dynamics and morphological evolution is an ongoing research topic. The kinetic Monte Carlo (KMC) method[1,2] that approximates a system evolution as succession of state-to-state Markov walks dictated by rates of all possible processes is taken as one of the method of choice in such studies. In KMC, the probability to execute a process is proportional to its rate and the time advanced after each process execution depends inversely on the sum of the rates of all possible processes in the system. The rate for atomic diffusion processes on surface follows Arrhenius rate expression from transition state theory[2,3]

$$k_{ij} = r_0 e^{\frac{-E_a}{k_B T}}, \tag{1}$$

where $k_{ij}$, $r_0$, $E_a$, $k_B$ and T are the rate of transition from state i to state j, an attempt frequency, activation energy barrier of the process, the Boltzmann constant, and the absolute temperature, respectively. The quantity $E_a$ is the energy difference between the maximum energy in the minimum energy path (MEP) and the energy of the stable initial configuration on which the process in question executes. For a given interatomic interaction, various methods[4-7] are in use to compute the MEP and then an $E_a$ of a process. Although such a calculation using a semi-empirical interatomic interaction is orders of magnitude faster in comparison to the first principles calculation based on density functional theory, it is still computationally intensive. Not surprisingly, attempts have been made to predict activation energy barriers from other considerations. Such studies can broadly be classified as methods that infer $E_a$ from values of other related quantities and those that predict $E_a$ directly. Along the first category, the diffusion barrier of atomic or molecular species is proposed to be 12% of the binding energy,[8] or the energy of end state,[9] or the energy difference between the final and initial configurations.[10] From chemical reaction studies Michaelides et al.[11] infer $E_a$s of dissociative reactions in heterogeneous catalysis using enthalpy changes, while in refs.[12-14] a linear correlation between the transition state energy and the final state energy is found by exploring the bond breaking reaction of diatomic molecules on surface catalysts. Similarly, Jacob et al.[15] have proposed oxygen p-band center as a measure of a perovskite compound's catalytic activity. In the second category, simplified approaches based on counting of broken and newly formed bonds,[16-19] and more sophisticated methods based on cluster expansion[20, 21], genetic programming[22], and artificial neural network[19, 23-27] are used to predict activation energy barriers. Both approaches provide convenient tools for computational screening of catalyst properties. Of course, their reliability depends on how accurately these predictions match real catalytic performance.

In island diffusion studies, although repeated calculation of barriers of single-atom, multi-atom, and concerted processes can be avoided by storing and retrieving the values from a database using a pattern recognition scheme, as exemplified in the self-learning kinetic Monte Carlo (SLKMC) method[28], calculation of barriers is a time consuming task. In this work we take advantage of the large database that has been acquired in previous applications of SLKMC by analyzing the dependence of the activation energy barriers on a variety of physically intuitive parameters. We then generate local geometric and energetic descriptors that might serve as predictors of these energy barriers for related systems. These descriptors are then used to develop predictive model to efficiently calculate barrier of processes for new systems.

In what follows, we provide computational details including considerations of descriptor generation and the basics of predictive models in section II. Results including comparison of predicted and calculated barriers are illustrated in section III, and conclusions are offered in section IV.

## II. COMPUTATIONAL DETAILS

### A. Database of energy barriers of diffusion processes

In this study information about the geometry of islands before and after execution of a process and the energy barriers of 844 processes are taken from our SLKMC study of homo-epitaxial adatom island (containing 2-4 atoms) diffusion on the (111) surfaces of Ni, Cu, Ag, and Pd. Some of the barriers are reported in our previous studies[29-32]. As discussed in those publications, the energy barriers were calculated using the embedded atom method (EAM)[33] interaction potentials for a system consisting the adatom island of interest and 5-layers of the substrate, with 256 (16x16) atoms per layer. Further details can be found in ref[28, 32]. Out of 844 diffusion processes considered here, 168 are taken from Cu/Cu(111), 191 from Ag/Ag(111), 156 from Pd/Pd(111) and 328 from Ni/Ni(111). The energy barriers for these processes range from 0.003eV to 1.302eV.

### B. Descriptor selection

As already mentioned, to identify descriptors we analyze the variation of the energy barrier of processes executed by a specific adatom of an island as it hops from one site to other allowed target sites on the underlying substrate, for a variety of configurations of the island. For example, we consider the initial and final geometry of the diffusing island, the local neighborhood of the adatom on the top substrate layer, and energetics such as the binding energy of the island with the substrate and lateral interactions within the island. The selection of proper descriptors and their sufficiency is very crucial to get high correlation between the predicted barriers, obtained from regression models, and those from interaction potentials. Below we provide the rationale used in the identification of descriptors using examples of processes found for the diffusion of Pd islands on the Pd(111) surface which have led to the identification of a set of predictors adequate for describing the diffusion energetics of our systems of interest here. Note that in the figures that follow, adatoms are represented by filled blue circles, the bonds between them by red lines, the executed processes by black lines with arrowhead, and the substrate mesh by orange lines at whose intersections (node) the substrate atoms sit.

*1. Number of bonds that change during diffusion ($x_1$)*

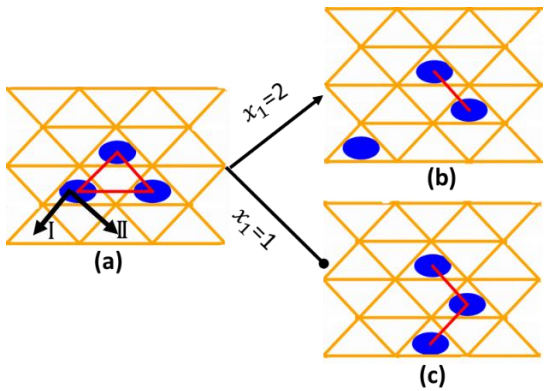

Figure 1. Illustration of descriptor $x_1$, the number of bonds change: (a) A trimer island with three bonds among adatoms in which the same adatom may undergoes two different single-atom diffusion processes to generate final structures shown on the right with one (b) and two (c) bonds, respectively.

In the trimer island shown in Fig.1a, the initial configuration has 3 bonds between the adatoms. On execution of either of the two processes indicated in the figure, the final configuration formed in Fig. 1 (b) has 1 bond and that in Fig. 1 (c) has 2 bonds, i.e. the change in the number of bonds, represented by the descriptor $x_1$, is 2 or 1, respectively. Here process with $x_1 = 2$ has a barrier of 1.147eV and that with $x_1 = 1$ has a barrier of 0.629eV, respectively. The higher activation energy barrier when larger number of bonds are broken is naturally not surprising, as it costs more energy to break two bonds than one..

*2. Shift of the island geometric center ($x_2$)*

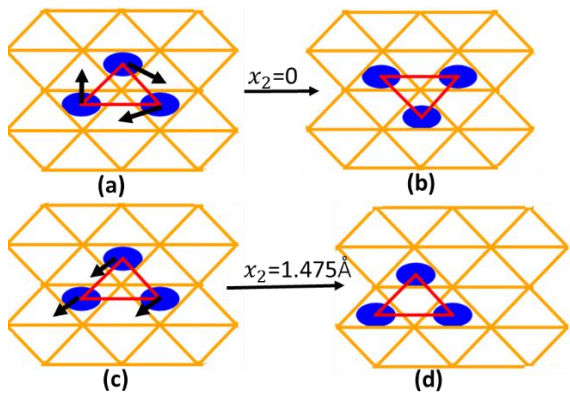

Figure 2. Illustration of descriptor $x_2$, shift of island geometric center: When the trimer island in (a) undergoes rotation to form the island shown in (b) $x_2= 0$. After executing the translation operation represented in (c), however, it takes the form shown in (d) for which $x_2= 1.475$Å.

For the trimer island shown in Fig. 2(a) or (c), the barriers of the shown processes: rotation from (a) to (b) and translation from (c) to (d) are 0.094eV and 0.151eV, respectively. Note that the value of the descriptor $x_1$ is zero for both processes, as no bonds are broken. Our choice of the descriptor $x_2$, the shift of the geometric center of the island, may serve as another measure of the differences in the energy barriers of processes in which there is no breaking of bonds, for the case shown in Fig. 2, The descriptor $x_2$ has values 0Å and 1.475Å for the rotational and translational process, respectively.

### 3. A-or B-type process ($x_3$)

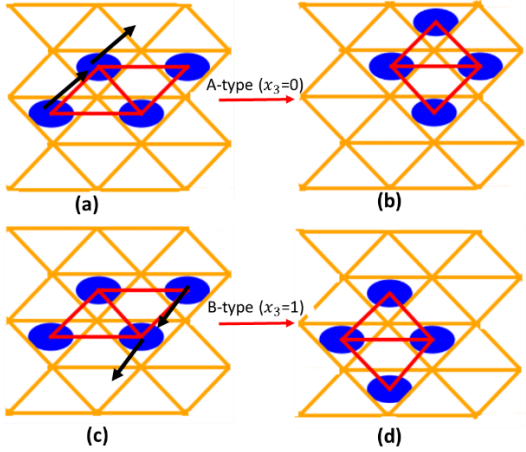

Figure 3. Illustration of descriptor $x_3$, the step micro-facets: Processes executed by two adatoms of an island in (a) and (c) with the same initial and final configurations as seen in (b) and (d) are distinct as the diffusion is along the (100) micro-facet or A-step in (a) and the (111) micro-facet or B-step in (c). We assign the descriptor $x_3$t values 0 or 1, as shown.

The edges of an island on fcc(111) surfaces form either the (100) micro-facet (called A-step) or (111) micro-facet (called B-step) with the underlying terrace. The two diffusive adatoms in the tetramer island shown with arrowhead in in Fig. 3 (a) form A-step while the other two atoms with arrows in Fig. 3(c)) form the B-step. For the two processes shown in Fig. 3(a) and 3(c), the tetramer islands have the same initial configurations and the same final configurations, as displayed in Fig. 3(b) and 3(d). The values of the descriptors $x_1$ and $x_2$ introduced above remain same for both processes. However, the barrier of the process executed by atoms in A-step (A-type process) is 0.319 eV and that of the process executed by atoms in B-step (B-type process) is 0.485eV, a significant variation. We thus introduce a binary descriptor $x_3$ that takes value 0 if the process is A-type and 1 if it is B-type.

*4. Number of diffusing atoms in a process ($x_4$)*

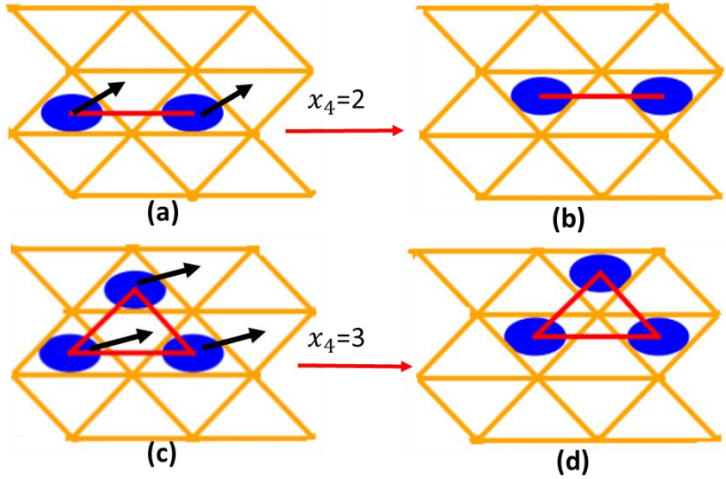

Figure 2. Illustration of descriptor $x_4$, the number of diffusing atoms: The translational concerted processes of a (a) dimer, and (c) trimer island are distinct on the number of adatoms involved in the process. The descriptor $x_4$ takes the value 2 and 3 for the processes shown in (a) and (c), respectively.

For small size of islands, concerted motion of an island is an important contributor to the diffusivity. In Fig. 4, concerted processes in a dimer and a trimer islands are shown for which the value of their activation energy barrier are 0.113eV and 0.155eV, respectively. Once again, for the processes considered in Fig. 4, the descriptors introduced earlier have the same value. The additional descriptor $x_4$ whose value represents the number of atoms in the island undergoing motion help account for the difference in the energy barriers as signified in Fig. 4.

*5. Binding energy of adatom island with substrate ($x_5$)*

Except for $x_3$, the shift of geometric center, the other descriptors discussed above cannot encode the variation of the barrier of the same process for different elements. For each cases we propose the binding energy of the island with the substrate to be an additional descriptor. To make our case, we first examine the variations in the activation energy barriers of small islands of several systems of interest.

The energy barriers in the first four columns in Table I show that those for the diffusion of Cu islands are smaller than that of corresponding process of Ni islands on either Ni(111) or Cu(111) substrate (see ref. 30 for detailed comparison of number of barriers of Cu and Ni islands on Ni(111)). Similar trend also holds when comparing Ag and Pd systems, as can be seen in last four columns in Table I in which the barriers of diffusion processes of Ag islands are in general larger than that of corresponding processes of Pd island on the Ag(111) or Pd(111) substrate. Although our training data in the work here is founded on homo-epitaxial system, we include hetero-epitaxial systems to show the general trend.

Table I. Comparison of energy barriers of diffusion processes of some selected single-atom (S), multi-atom (M), and concerted (C) processes of adatom islands on several homo-epitaxial and hetero-epitaxial islands on fcc(111) system.

| Island Size | Energy barrier(eV) | | | | | | | |
|---|---|---|---|---|---|---|---|---|
| | Cu/Ni | Ni/Ni | Cu/Cu | Ni/Cu | Pd/Ag | Ag/Ag | Pd/Pd | Ag/Pd |
| 1 | 0.052 | 0.059 | 0.030 | 0.032 | 0.054 | 0.059 | 0.037 | 0.055 |
| 2(S) | 0.029 | 0.034 | 0.013 | 0.028 | 0.033 | 0.055 | 0.026 | 0.065 |
| 2(C) | 0.059 | 0.066 | 0.019 | 0.021 | 0.066 | 0.098 | 0.043 | 0.065 |
| 3(C) | 0.152 | 0.187 | 0.103 | 0.132 | 0.241 | 0.176 | 0.168 | 0.176 |
| 4(M) | 0.191 | 0.271 | 0.194 | 0.282 | 0.355 | 0.244 | 0.320 | 0.221 |

Table II. Size-dependent adatom island binding energy for several homo-epitaxial and hetero-epitaxial islands on fcc(111) substrate.

| Island Size | Island binding energy (eV) | | | | | | | |
|---|---|---|---|---|---|---|---|---|
| | Cu/Ni | Ni/Ni | Cu/Cu | Ni/Cu | Pd/Ag | Ag/Ag | Pd/Pd | Ag/Pd |
| 2 | -3.29 | -3.28 | -3.04 | -3.04 | -3.94 | -1.96 | -3.89 | -2.93 |
| 3 | -4.24 | -4.18 | -3.89 | -3.83 | -5.04 | -2.49 | -5.07 | -3.84 |
| 4 | -5.37 | -5.29 | -4.93 | -4.86 | -6.33 | -3.16 | -6.40 | -4.93 |

To obtain qualitative understanding of the above trends in the diffusion barriers presented in Table I, we present in Table II the binding energy (B.E.) of adatom islands on the respective surface mentioned in the Table I. These are calculated using the expression

$$B.E. = E_{island+subs.} - E_{island} - E_{subs.}, \qquad (2)$$

where $E_{island+subs.}$, $E_{island}$, $E_{subs.}$ are the total energy of a system containing the adatom island on the substrate, the isolated island, and the isolated substrate, respectively. One can see from Table II that the B.E. of Cu islands on Ni(111) are larger than that of Ni islands on the same surface, while that of Pd islands are larger than that of Ag islands on the same substrate. We conclude that the activation energy barriers of processes correlate inversely with the island binding energy with the substrate. Based on the above analysis we suggest the binding energy of the adatom island as the fifth descriptor $x_5$.

### 6. Lateral interaction energy among adatoms ($x_6$)

In an earlier work[34], we had emphasized the role of lateral interaction to understand the relatively smaller barriers of the diffusion processes of Cu adatom islands than that of Ni on the Ni(111) substrate. The slight difference in the binding energy of Cu and Ni islands on Ni(111) (first 2 columns of Table II) also indicates that several factors beyond binding energy are responsible for the noticeable differences in diffusion barriers for otherwise similar processes. To examine the role of lateral interactions in a general manner as implicated in the differences in barriers of multi-atom and concerted processes, we turn here to a comparison of some frequently-executed multi-atom and concerted processes of islands of different sizes for four systems in Table III.

Table III. Comparison of barriers of concerted and multi-atom processes for the same adatom island configuration in homo and hetero-epitaxial systems.

| Island Size | Energy barrier (eV) | | | | | | | |
|---|---|---|---|---|---|---|---|---|
| | Pd/Pd | | Ag/Ag | | Cu/Ni | | Ni/Cu | |
| | Concerted | Multi-Atom | Concerted | Multi-Atom | Concerted | Multi-Atom | Concerted | Multi-Atom |
| 4 | 0.186 | 0.324 | 0.190 | 0.246 | 0.182 | 0.205 | 0.157 | 0.303 |
| 5 | 0.277 | 0.345 | 0.281 | 0.254 | 0.235 | 0.196 | 0.220 | 0.313 |
| 6 | 0.284 | 0.643 | 0.299 | 0.551 | 0.290 | 0.580 | 0.222 | 0.731 |
| 7 | 0.327 | 0.418 | 0.417 | 0.319 | 0.460 | 0.318 | 0.431 | 0.277 |
| 8 | 0.416 | 0.665 | 0.401 | 0.469 | 0.378 | 0.477 | 0.356 | 0.611 |

In Table III, one can make two observations: there is a noticeable difference in the barriers of concerted and the multi-atom processes for the same island configuration and that the magnitude of the difference is system dependent. We find the order to be Ni/Cu, Pd/Pd, Cu/Ni, and Ag/Ag i.e, the difference is in general large in Ni/Cu and Pd/Pd systems. Such a difference might be understood from the interatomic interaction among adatoms. One quantitative measure of such is the lateral interaction amongst the adatoms in the island which is defined by

$$E_{L.I.} = E_{island+sub.} - nE_{mono+sub.} + (n-1)E_{sub.} \qquad (3)$$

where $E_{island+sub.}$, $E_{mono+sub.}$, and $E_{sub.}$ are the total energy of the system with the island on the substrate, a monomer on the substrate, and that of an isolated substrate, respectively. The calculated values of $E_{L.I.}$ for four different systems whose energy barriers are in Table III are presented in Table IV. Note again that although our training data is for homo-epitaxial systems, we have brought hetero-epitaxial systems to clarify the ideas.

Table IV. Lateral interaction among adatoms of islands on the fcc(111) substrate for configurations relevant to processes in Table III.

| Island Size | Lateral interaction energy (eV) | | | |
| --- | --- | --- | --- | --- |
| | Pd/Pd | Ag/Ag | Cu/Ni | Ni/Cu |
| 2 | -0.54 | -0.32 | -0.39 | -0.54 |
| 3 | -1.49 | -0.89 | -1.09 | -1.44 |
| 4 | -2.42 | -1.42 | -1.89 | -2.30 |
| 5 | -3.29 | -1.94 | -2.58 | -3.12 |
| 6 | -4.2 | -2.46 | -3.28 | -3.96 |
| 7 | -5.45 | -3.19 | -4.26 | -5.13 |
| 8 | -6.33 | -3.69 | -4.94 | -5.94 |

From Table IV, one can see that the Pd/Pd and Ni/Cu systems which have the relatively large difference between barriers of concerted and multi-atom processes, the former having lower than the latter, have strong lateral interactions among adatoms whereas for the Ag/Ag and Cu/Ni systems such an interaction among adatoms is weak. We thus propose the among adatoms as out sixth $x_6$.

C. Data distribution and correlation among descriptors

The frequency distribution of the values of each of the descriptors and activation energy barriers in our database along diagonal boxes in the scatter plot is shown in Fig. 5. The off-diagonal boxes in the figure show the pairwise distribution of descriptors and their correlation measured as Pearson's correlation coefficient using origin lab software (Origin 2016, OriginLab Corporation, Northampton, USA).

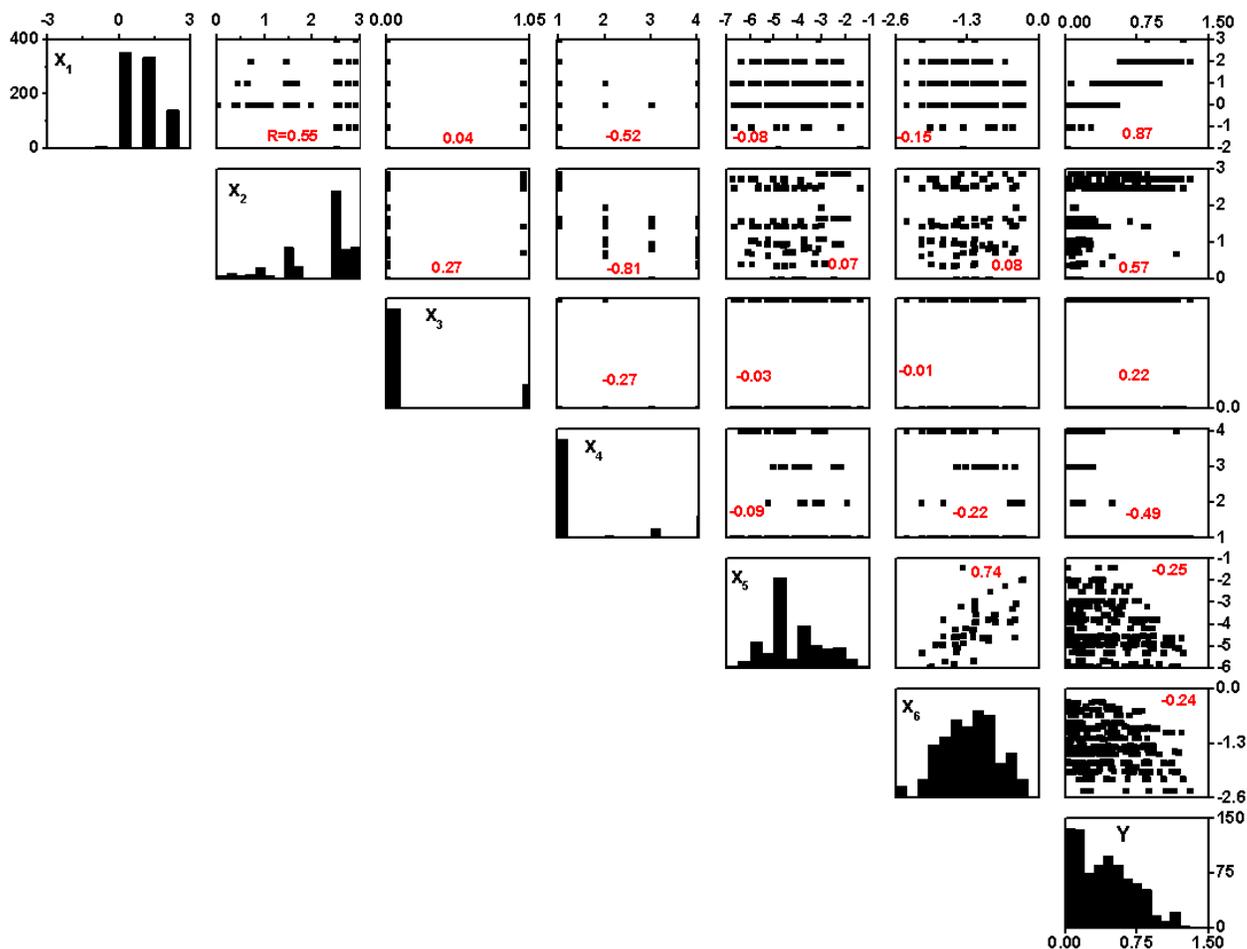

Figure 5. A scatter matrix plot that shows the frequency distribution of the values of descriptors used in this study along diagonal boxes and the pairwise correlation between variables on off-diagonal boxes. The values of Pearson's correlation coefficient for each pair are shown in each off-diagonal boxes.

From the frequency distribution of descriptor $x_1$ (number of bond changes) in our database (the top left box in Fig. 5), one can see that it has positively skewed distribution. The number of instances with bond change of -2, -1, 0, 1, 2, and 3 are 2, 10, 352, 335, 141 and 4, respectively. So, there are large number of processes in our database in which either the number of bonds does not change or decrease by 1 or 2 after executing them. The value of the correlation coefficient between the descriptor $x_1$ and other descriptors are shown on five successive boxes on the top row. The strong positive correlation coefficient of $x_1$ with $x_2$(shift of island geometric center) means that in number of instances in which the number of bond changes increases, the shift of island geometric center increases. The negative correlation with $x_4$(number of diffusion atoms) indicate the small value of change in the number of bonds has large number of atoms involved in diffusion and vice versa, for e.g., concerted process with zero value of bond change has more atoms involved than single atom detachment process that has positive value of bond change with one number of atom involved. Lack of definite tendency between $x_1$ with $x_3$(A- or B-type process), $x_5$ (island binding energy on substrate) and $x_6$ (lateral interaction energy among adatoms) as seen by small value of correlation coefficient indicates that they are independent. The descriptor $x_1$ has high positive correlation with activation energy barrier with value of Pearson's coefficient being 0.87, the highest among the selected descriptors which reflects that processes with relatively large number of bonds change have relatively higher barrier.

The descriptor $x_2$ has values that distribute from 0 to 3Å. It has strong negative correlation with $x_4$ which reflects the fact that the most of single atom processes are long jump with high value of distance travelled whereas the multi-atom or concerted are short jump. The Pearson's coefficient between $x_2$ and activation energy barrier is 0.57, which reflects the fact that long jump processes that shift the center more have relatively higher barriers.

The distribution of values of $x_3$ shows that there are large number of A-type processes in our database. The small negative correlation between $x_3$ and $x_4$ indicates that in our database A-type processes have more number of diffusion atoms than B-type processes. The descriptor has positive correlation (coefficient = 0.22) with activation energy barrier reflects that B-type processes have relatively high barrier. The descriptor has negligible relation with other descriptors.

The distribution of values of descriptor $x_4$ indicates that there are large number of single atom processes. The negative correlation with $x_5$ reflects that there are the single atom processes or less number of atom moving processes most in configurations or systems with higher binding energy with substrate. The negative correlation with activation energy barrier reflects the relatively small value of barriers of multi-atom or concerted processes in comparison to that of many detaching single atom processes in our database.

The values of descriptor $x_5$ has values in range -7eV to 1 eV and has negative correlation with the barrier of processes. It has strong correlation with value 0.75 with descriptor $x_6$ indicating that the system with larger binding energy with substrate also has larger lateral interaction. Such a tendency can also be seen from Table I and Table IV in which Pd/Pd system has both quantity the largest and Ag/Ag has both the lowest but it does not hold in Cu/Ni and Ni/Cu system. The descriptor $x_6$ (lateral interaction energy) has almost normally distributed values. It has negative correlation with descriptor $x_4$ implying structures with higher lateral interaction has small barrier for processes with higher number of atoms, in agreement with the small barrier of concerted processes in comparison to that of multi-atom process in system with larger lateral interaction which is already discussed when describing descriptor $x_6$. The values of these six descriptors and the barrier of processes are used in a data-enabled model to predict the barriers of diffusion processes.

### D. Predictive models

#### 1. Multivariate linear regression (MLR)

As an initial step, a predictive model based on multivariate linear regression technique which assumes independent variables linearly contribute to the dependent variable is tested whose mathematical form is

$$y_i = \beta_0 + \sum_{j=1}^{6} \beta_j x_j + \epsilon_i, \tag{4}$$

where $y_i$ represents the predicted value of dependent variable ($E_a$ of a diffusion process in this study) on the $i^{th}$ instance ($i$ has value 844 when all data is used to develop model in this study) and $x_j$ is the value of $j^{th}$ independent variables, (j=1, 2, ..6) on the same instance. The parameters $\beta_0$ and $\beta_j$s represent an intercept and the predictor $x_j's$ slope, respectively and $\epsilon_i$ represents error term on $i^{th}$ instance. Taking error function as the sum of square of errors and applying the condition of its minimization with respect to parameter βs, one can get a matrix of values of fitting parameters βs using

$$\beta = (X^T X)^{-1} X^T Y, \tag{5}$$

where X refers a matrix formed with values of 6 descriptors along column and training samples as rows, T refers the transpose operation of matrix and Y is a column matrix formed by calculated activation energy barriers, $c_i$. The Pearson's correlation coefficient (R) between $c_i$s and $y_i$s is used for quantitative measure of the predictive capacity of the model which is calculated using

$$R = \frac{N \sum_i y_i c_i - \sum_i y_i \sum_i c_i}{\sqrt{[N \sum_i y_i^2 - (\sum_i y_i)^2][N \sum_i c_i^2 - (\sum_i c_i)^2]}}, \tag{6}$$

where N represents the sample size used to predict.

*2. Non-linear model using Artificial Neural Network (ANN)*

In this study, the possibility of a non-linear dependence between independent and dependent variables is modeled by using artificial neural network (ANN)[35] approach as implemented in MATLAB software (MATLAB and Statistics Toolbox Release 2017b, The MathWorks, Inc., Natick, MA, USA). Because of the ability of ANN to identify underlying highly complex and non-linear relationships on input-output data, it become a method of our choice for data fitting purposes. The structural organization of the ANN used in this study consists of input with 6 nodes corresponding to 6 input features ($x_j s$), 2 hidden layers containing 10 nodes each with sigmoid and linear transfer function for 1st and 2nd hidden layers respectively, and output layer. Following the fitting mechanism of an ANN, the mathematical expression for the value on $k^{th}$ node (k=1, 2, ..,10) in 1st hidden layer ($x_k^1$) is given by

$$f\{\sum_{j=1}^{6}(w_{kj}^1 x_j + b_k^1)\} = x_k^1,$$

where the superscript refers hidden layer number, $w_{kj}^1$ and $b_k^1$ are the weight and bias at node k in 1st hidden layer for each instances of input $x_j$ in input layer and f is transfer function (sigmoid in this study). The value on $p^{th}$ node (p=1,2, ..,10) in 2nd hidden layer ($x_p^2$) is given by

$$f\{\sum_{k=1}^{10}(w_{pk}^2 x_k^1 + b_p^2)\} = x_p^2,$$

where the second transfer function is linear function in this study. The output ($y_i$) which gives the predicted barrier is given by

$$y_i = f\{\sum_{p=1}^{10}(w_{ip} x_p^2 + b_i)\}. (7)$$

The mean square error (MSE) is taken as error function which is calculated as

$$\text{Error function} = \text{MSE} = \sum_{l=1}^{N} \frac{(y_i - c_i)^2}{N},$$

where $c_i$ refers to the calculated energy barrier or target value assigned at the beginning of fitting and N is the number of training samples. During the training process, the weights and biases are adjusted to optimize the function that converts the neural network training problem into an optimization problem. Levenberg-Marquardt_algorithm[36, 37] is used as optimizer in which the current weight and bias vector ($v_k$) is updated to vector $v_{k+1}$ following

$$v_{k+1} = v_k - (JJ + I * \mu)J_e,$$

where I is an identity matrix, μ is a training rate parameter which influences the rate of weight and bias adjustment and $JJ = (J_F)^T (J_F)$ for $J_F$ being the Jacobian of performance function with respect to weights and biases.. In this study, the initial weight and biases are randomly selected and the initial value of the learning rate parameter µ is taken as 0.001 which is increased by a factor of 10 until the updated $x_{k+1}$ results in a reduced performance after which µ is decreased by a factor of 0.1. During training, the network performance in the validation vectors is checked every epoch and if it increases or remains constant in the current step for 6 additional epochs in a row or the performance error is 0 or the gradient of the error is less than $10^{-7}$, the calculation is terminated.

## III. RESULTS

In this section, we present results of predicted diffusion barriers obtained from using multivariate linear regression and neural network techniques.

In the first set of calculation, a multivariate linear model is formed by taking each atomic system separately. The value of $R^2$ obtained using Equation (6) and the expansion coefficients as obtained by using equation (5) are presented in Table V. A model formed by taking all 844 samples of the input-output pair is used to predict values of Ni barriers in which case the value of $R^2$ between the calculated and the predicted values is 0.893 (R =0.944). The high value of $R^2$ indicates that the 6 descriptors used in this study are sufficient to explain the variability of output variable (activation energy barrier) and can predict barrier reliably.

Table V. Calculated coefficients in the linear predictive equation of activation energy barrier using 6 descriptors and value of $R^2$ for models.

| System | Sample Size | Coefficients | | | | | | | $R^2$ |
|---|---|---|---|---|---|---|---|---|---|
| | | $\beta_1$ | $\beta_2$ | $\beta_3$ | $\beta_4$ | $\beta_5$ | $\beta_6$ | Intercept | |
| Cu | 168 | 0.327 | -0.09 | 0.098 | 0.062 | 0.029 | 0.002 | -0.094 | 0.96 |
| Ag | 191 | 0.232 | -0.057 | 0.107 | 0.055 | 0.223 | -0.240 | 0.274 | 0.904 |
| Pd | 157 | 0.472 | 0.119 | 0.203 | 0.057 | -0.007 | 0.048 | -0.157 | 0.95 |
| Ni | 328 | 0.323 | 0.111 | 0.132 | 0.056 | 0.016 | -0.019 | -0.63 | 0.922 |
| CuAgPdNi | 844 | 0.314 | 0.085 | 0.119 | 0.036 | -0.058 | 0.041 | -0.277 | 0.841 |
| Ni using CuAgPdNi | 328 | 0.898 | | | | | | 0.0303 | 0.893 |
| CuAgPd | 516 | 0.318 | 0.095 | 0.109 | 0.047 | -0.094 | 0.122 | -0.365 | 0.814 |
| Ni using CuAgPd | 328 | 0.889 | | | | | | 0.036 | 0.85 |

As a second scenario, we use the dataset of Cu, Ag, and Pd systems that contains 516 samples to develop a model and test its predictive capacity for the Ni system. The value of $R^2$ in this case comes out to be 0.85 (R= 0.921). The calculated and predicted activation energy barriers of the Ni system in the first and second scenario are plotted in Fig. 6(a) & (b), respectively. The blue continuous line represents an ideal case.

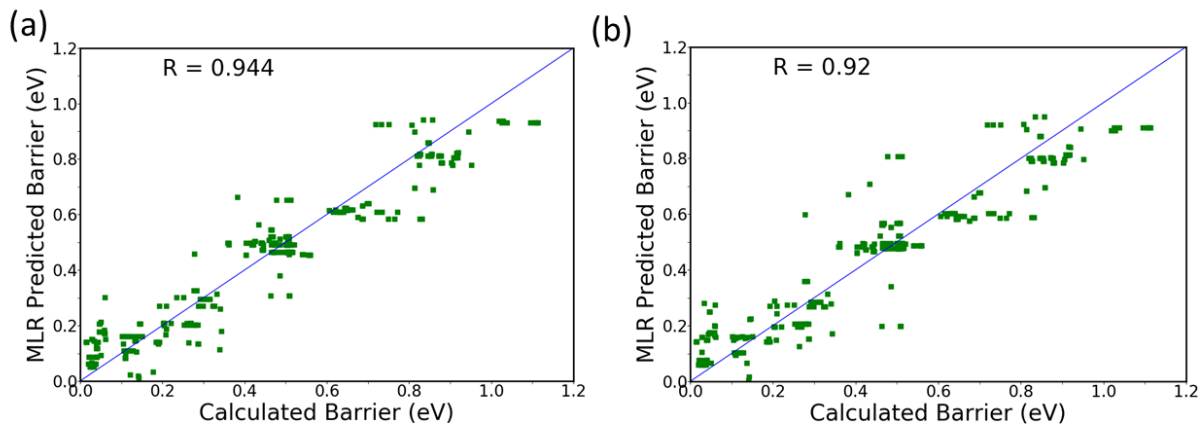

Figure 3. Predicted vs. calculated barriers of processes for Ni islands on the Ni(111) surface, (a) when some Ni samples are used to train of linear model and (b) Ni samples are not used to train the model.

When using neural network, training, validation, and testing data sets are randomly selected from input data in 70%, 15% and 15%, respectively. The value of correlation coefficients between calculated and predicted barriers are presented in Table VI.

Table VI. Values of correlation coefficients in the training, validation, and testing dataset of diffusion barriers using the neural network.

| System | Sample Size | Correlation Coefficient (R) | | |
|---|---|---|---|---|
| | | Training | Validation | Testing |
| Cu | 168 | 0.99 | 0.98 | 0.98 |
| Ag | 191 | 0.99 | 0.99 | 0.99 |
| Pd | 157 | 0.99 | 0.97 | 0.98 |
| Ni | 328 | 0.99 | 0.98 | 0.99 |
| CuAgPdNi | 844 | 0.97 | 0.97 | 0.97 |

The plots of the predicted vs. computed values of barriers of processes for training, validation, and testing data are shown in Fig. 7(a), (b), and (c) respectively. The high value of correlation coefficient has demonstrated that artificial neural network model is good for predicting activation energy barriers.

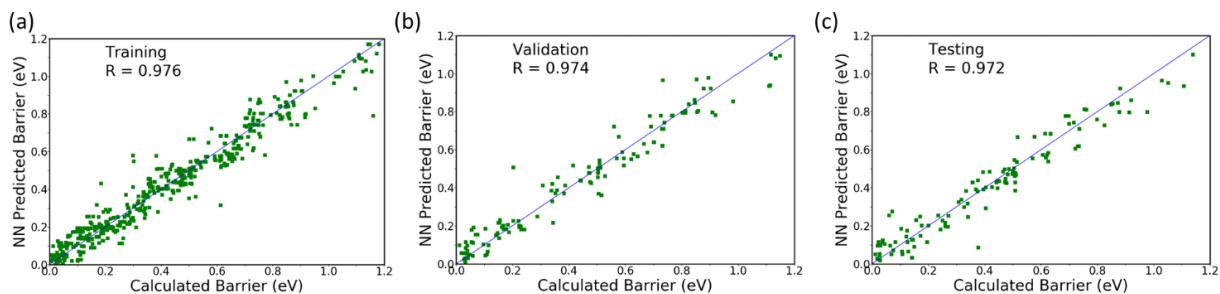

Figure 4. Calculated vs. predicted barriers of different processes of an island diffusion on the fcc(111) surface using a neural network; (a), (b), and (c) show plots for training, validation, and testing samples, respectively.

To test the generality of the model, we predict the activation energy barriers of processes for the Ni excluding the barriers of the system in the training and validation process. A plot of the predicted vs calculated barriers is shown in Fig. 8 and a histogram plot of the error of predictions is shown in Fig. 9.

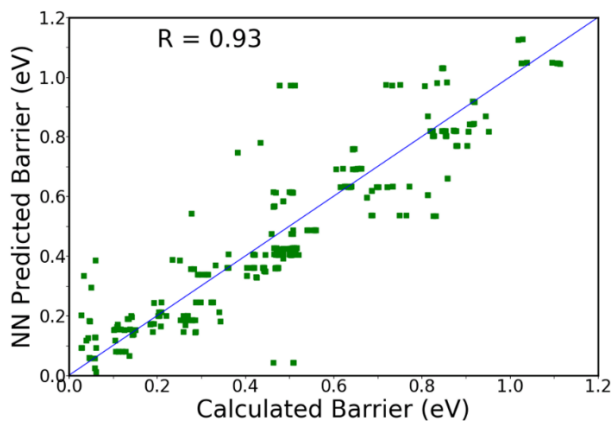

Figure 8. Calculated vs. predicted barriers of different processes of Ni island diffusion on the Ni(111) surface using a neural network model without using Ni barriers in training and validation process.

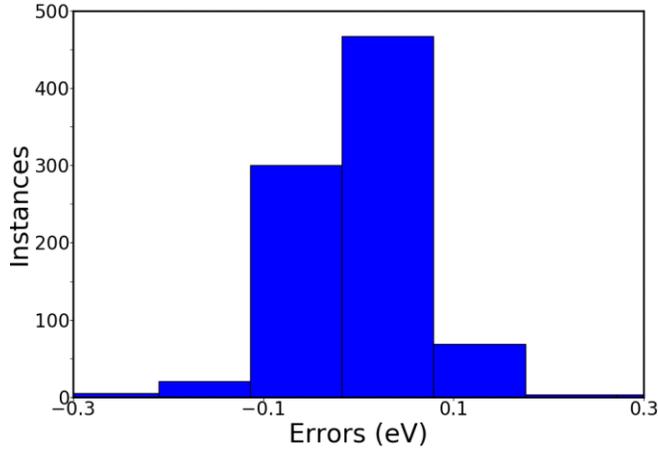

Figure 9. Error histogram of prediction of barriers of Ni/Ni(111) processes using a neural network testing process without using Ni barriers in training and validation process.

From Fig. 9, one can see that there are large number of instances where the error lies within few meV of the calculated values and the instances with increasing error magnitude decrease with increase in error value.

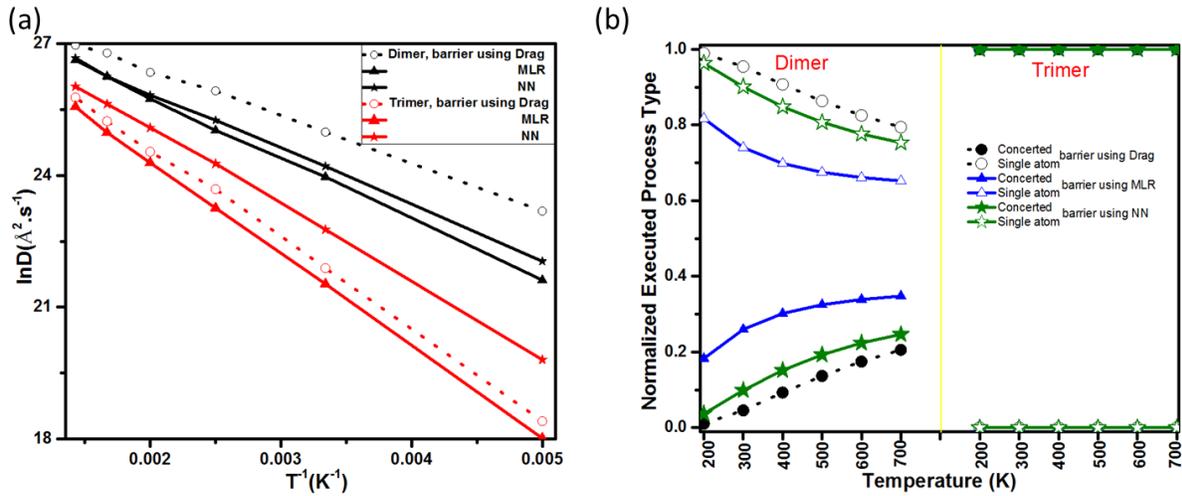

Figure 10. (a) Arrhenius plot, (b) normalized type of executed processes obtained from KMC simulation of the Ni dimer and trimer diffusion processes on the Ni(111) surface using barriers from drag method from EAM interaction and from the trained neural network model.

The predicted barriers are then used to explore the diffusion properties of the island using the kinetic Monte Carlo method. The Arrhenius plot and the normalized type of executed processes (single, multi-atom, concerted) in the simulation of dimer and trimer island diffusion in the temperature range 200K to 600K are presented in Fig. 10.

The close agreement of the quantities plotted in Fig. 10 imply that the predictive approach can be used to find reliable diffusion characteristics of islands of any size of any element in the set of metallic systems considered here, with a significant gain in computational time.

## IV. CONCLUSIONS

In this work we have applied a multivariate linear and nonlinear regression models to train and thereby predict the activation energy barriers for the diffusion of small, two-dimensional adatom islands containing 1-4 transition metal atoms, on their respective (111) surfaces. The training set was extracted from a large database of activation energy barriers, calculated using semi-empirical interaction potentials, for diffusion processes (single and multiple atoms) revealed using a Self-Learning Kinetic Monte Carlo scheme. A comprehensive analysis of this large set of barriers led the identification of 6 descriptors, 4 geometric and 2 energetic, which explain the observed trends. The values of these descriptors and activation energy barrier of diffusion processes are then used to train, validate, and test models in linear approximation using multivariate linear regression and non-linear approximation using the neural network approach. After parametrization, the models are used for ultrafast prediction of activation energy barriers for those systems in both supervised and unsupervised mode. The high value of correlation coefficients of predicted barriers with computed ones using energy minimization techniques make this approach very promising for application of tools suitable for multi-scale modeling of thin film growth and morphological evolution of nanostructured systems. As expected, the comparison of results of diffusion properties in a test case of Ni dimer and trimer island diffusion on the Ni(111) surface obtained from KMC simulation using such predicted barriers captures the features from such study using barriers from interatomic interaction potential. While this study was focused on homo-epitaxial systems consisting of transition metal with fcc stacking, ther results and the approach is quiet general and should be applicable to a variety of systems of interest.

ACKNOWLEDGMENTS

We would like to acknowledge the computational resources provided by the STOKES facility at the University of Central Florida. We thank Duy Le and Mikko Hakala for suggestions for improvements of the manuscript. We are pleased to acknowledge NSF for partial support under grant DMR-1710306.